\documentclass[sigconf]{acmart}
\usepackage{tabularx}
\usepackage{algorithm}
\usepackage{algpseudocode}
\usepackage{listings}
\usepackage{enumitem}

\AtBeginDocument{%
  }

\copyrightyear{2026}
\acmYear{2026}
\setcopyright{cc}
\setcctype{by}
\acmConference[ICPP Workshops '26]{Workshop Proceedings of the 55th International Conference on Parallel Processing}{September 28-October 01, 2026}{Singapore, Singapore}
\acmBooktitle{Workshop Proceedings of the 55th International Conference on Parallel Processing (ICPP Workshops '26), September 28-October 01, 2026, Singapore, Singapore}
\acmDOI{10.1145/3816891.3834899}
\acmISBN{979-8-4007-2763-4/2026/09}

\begin{document}

\title{A Preliminary Study on Simultaneous Coscheduling \\ for Discrete GPU vs.\ Fused GPU}

\settopmatter{authorsperrow=4}
\author{Poorna Gunathilaka}
\affiliation{%
  \institution{Virginia Tech}
  \city{Blacksburg}
  \country{USA}
}
\email{poornag@vt.edu}

\author{Nabayan Chaudhury}
\affiliation{%
  \institution{Virginia Tech}
  \city{Blacksburg}
  \country{USA}}
\email{nabayanc@vt.edu}

\author{Kirshanthan Sundararajah}
\affiliation{%
  \institution{Virginia Tech}
  \city{Blacksburg}
  \country{USA}
}
\email{kirshanthans@vt.edu}

\author{Wu-chun Feng}
\affiliation{%
  \institution{Virginia Tech}
  \city{Blacksburg}
  \country{USA}
}
\email{wfeng@vt.edu}

\begin{abstract}
CPU-GPU coscheduling enables simultaneous execution of an application across both processing units, but its efficiency depends on workload partitioning and memory architecture. This preliminary study evaluates coscheduling on the NVIDIA GH200 Superchip compared to a discrete H100 PCIe platform. Using sparse conjugate gradient (CG) as a case study, we assess various work divisions across three memory-management paradigms: explicit copy, managed memory, and mapped memory. Our evaluation highlights the run time and programmability tradeoffs of reducing
manual CPU--GPU data movement.
The results show that compared with the H100 PCIe platform, GH200 makes several hybrid CPU--GPU work divisions competitive and makes managed memory practical for several matrices. These results suggest that integrated CPU--GPU platforms such as GH200 can improve both performance and 
programmability for coscheduled workloads.
\end{abstract}


\maketitle


\section{Introduction}

GPU-accelerated systems are commonly programmed using a discrete accelerator
model in which the CPU launches GPU kernels and manages data movement between
host and device memory~\cite{li2024automaticblas}. Although effective for many
applications, this model often separates CPU execution, GPU execution, and data
movement into distinct phases, leaving the CPU primarily responsible for
coordination.

CPU-GPU coscheduling provides a different execution model. In this paper, we
use \emph{CPU-GPU coscheduling} to refer to simultaneous execution in which CPU
cores and GPU cores compute different portions of the same application.  Its effectiveness
depends on workload partitioning, synchronization, and communication
costs~\cite{scogland2014coretsar}. Consequently, the best configuration depends
not only on GPU capability, but also on the workload, input, partitioning,
memory-management strategy, and platform architecture.

These factors are particularly important when comparing discrete GPUs with
integrated platforms such as the NVIDIA Grace Hopper (GH200) Superchip, which
connects a Grace CPU and Hopper GPU through the coherent, high-bandwidth
NVLink-C2C interconnect~\cite{wei2023nvlinkc2c}. However, tighter integration does
not eliminate memory-management concerns. GH200 performance remains sensitive
to data placement, migration, remote access, and coherence
behavior~\cite{schieffer2024harnessing,fusco2024understanding,werner2025disaggregation,bagchi2026consistency}.
Thus, integration can reduce CPU-GPU cooperation costs, but appropriate workload
partitioning and memory strategies remain necessary.

The choice of memory-management strategy also affects both performance and
programmability. Explicit-copy implementations provide
direct control over data movement but require the programmer to manually manage
transfers and shared-state exchange. Managed and mapped~\cite{cuda_programming_guide} memory can
reduce this programming burden by simplifying CPU-GPU data sharing, although
their run time cost depends on the access pattern and platform~\cite{schieffer2024harnessing,fujita2025sharedmemory,li2024automaticblas}.
Therefore, an integrated platform is useful not only if it improves run time, but also if it enables simpler memory-management strategies that are practical for coscheduled execution.

To study these tradeoffs we use sparse
CG, whose repeated shared-vector accesses and adjustable CPU-GPU work
division expose the effects of partitioning and memory management.

This paper makes the following contributions:
\begin{itemize}[itemsep=2pt,topsep=3pt,leftmargin=*]
    \item Characterization of the run time configuration landscape of sparse CG on
    an integrated NVIDIA GH200 
    and a discrete H100.

    \item Quantification of how CPU-GPU work division and memory-management strategy
    affect the practicality of coscheduled execution on both discrete and fused GPUs. 

    \item Evaluation of the run time and programmability tradeoff between
    explicit-copy execution, managed memory, and mapped memory.

    \item 
    An optimistic interconnect-bound analysis for the NVIDIA H100 (PCIe) by
    scaling measured explicit transfer time using an NVLink-like
    peak-bandwidth ratio.
    
\end{itemize}
Our results show that the NVIDIA GH200 Superchip broadens the practical sparse CG coscheduling space: 
Compared to the H100 GPU (PCIe), the GH200 makes mixed CPU-GPU work divisions more
competitive and makes managed memory feasible in many cases. 
Our
interconnect-bound analysis further shows that this advantage is not explained
by PCIe bandwidth alone.
\section{Background}
\label{sec:background}

This section outlines the execution-cost model, memory-management mechanisms, and architectural characteristics relevant to CPU--GPU coscheduling on integrated and discrete platforms.

\subsection{Run time Costs in CPU-GPU Coscheduling}
\label{sec:background-coscheduling-costs}

CPU-GPU coscheduling replaces the accelerator-offloading execution model with cooperative execution:
the CPU and GPU process different work partitions concurrently and synchronize
at algorithmic boundaries. Prior
coscheduling systems~\cite{scogland2014coretsar} model the
run time of such execution as the run time of the slower partition plus the costs of memory
movement and the synchronization required to make the two partitions operate
as a single computation. 
\vspace{-1mm}
\[
    T_{\text{cosched}} = \max(T_{\text{CPU}}, T_{\text{GPU}}) + T_{\text{mem}} + T_{\text{sync}}.
\]
\vspace{-4mm}

Here, \(T_{\text{CPU}}\) and \(T_{\text{GPU}}\) represent the execution time of
the CPU and GPU portions of the work. The memory term, \(T_{\text{mem}}\),
captures the cost of moving or sharing data between CPU and GPU memory domains,
including explicit transfers, shared-state exchange, migration, or remote
access. The synchronization term, \(T_{\text{sync}}\), captures barriers,
reductions, and other coordination needed between CPU and GPU execution.

This model highlights why coscheduling is platform-dependent. A CPU-GPU split
is useful only when the benefit of using both processors is larger than the
additional memory and synchronization cost. On a PCIe-attached GPU, repeated
shared-state exchange can exacerbate these costs. On an integrated platform such
as GH200, the CPU-GPU interconnect and memory system may reduce these costs,
making a wider range of workload partitions practical.

\subsection{Memory Management and Programmability}
\label{sec:background-memory}

Memory management affects both the performance and programmability of
CPU-GPU coscheduling. In explicit-copy execution, the programmer controls when
data moves between host and device memory. This gives precise control over data
placement but requires manual transfers and synchronization whenever shared
state must be exchanged.

Managed memory and mapped memory provide simpler alternatives. Managed memory
exposes a shared logical allocation whose placement and movement are handled by
the CUDA runtime~\cite{cuda_programming_guide}. Mapped memory exposes
page-locked host memory in the GPU address space, allowing GPU kernels to access
host-resident data without an explicit copy~\cite{cuda_programming_guide}. Both
approaches can reduce the programming burden associated with explicit data movement, but they do not
remove the cost of sharing data. Their performance still depends on access
pattern, data placement, migration, remote access, and platform support.

Therefore, these strategies represent a performance and programmability tradeoff.
Explicit copy provides the most control with higher programming effort,
whereas managed and mapped memory simplify data sharing but may introduce run  time
overhead. A central question in this paper is whether GH200 makes these
simpler strategies practical enough for coscheduled execution.

\subsection{Related Work}

\label{sec:bg-integrated-platforms}

Discrete GPU systems maintain separate CPU and GPU memory domains connected
through PCIe, making CPU-GPU cooperation dependent on transfers, synchronization,
or remote access. These costs can limit coscheduling when shared state is
exchanged frequently.

The NVIDIA GH200 
Superchip provides a different hardware setting by
connecting a Grace CPU and Hopper GPU using NVLink-C2C. Prior work describes
NVLink-C2C as a coherent chip-to-chip interconnect designed to provide tighter
CPU-GPU coupling than conventional PCIe-attached GPU systems~\cite{wei2023nvlinkc2c}. 
However, integration does not make memory behavior uniform. Schieffer et
al.~\cite{schieffer2024harnessing} show that system-allocated memory and CUDA
managed memory on the GH200 remain sensitive to page placement,
initialization, and migration. Werner et
al.~\cite{werner2025disaggregation} show that coherent CPU access to GPU memory
over NVLink-C2C can be useful but is not equivalent to local memory access.
Recent work also studies the consistency and coherence behavior of the
GH200 memory system~\cite{bagchi2026consistency}.

Application studies similarly show that GH200 simplifies data sharing but does
not eliminate data-movement costs. Shared-memory and BLAS studies report
tradeoffs among explicit copies, unified access, migration, and peak
performance~\cite{fujita2025sharedmemory,li2024automaticblas}, while broader
evaluations assess Grace-Hopper for HPC workloads~\cite{banchelli2024grace}.
Together, these results show that GH200 expands the memory-sharing design space,
but performance still depends on data placement, access direction, and workload
behavior.
This paper complements prior studies on GH200 characterization and programming
models by focusing on application-level CPU--GPU coscheduling. Rather than
studying only memory bandwidth, page migration, or a single programming model,
we evaluate how workload partitioning and memory-management strategy jointly shape
the run time on integrated and discrete GPU platforms.

\section{Design and Implementation}
\label{sec:design}

This section describes the hybrid CPU-GPU CG solver used as our case study application. CG solves a sparse
linear system \(Ax=b\), where \(A\) is fixed during the solve and \(x\)
is updated iteratively. Sparse CG is a useful target for this study
because sparse computations often expose irregular memory accesses,
low arithmetic intensity, and load imbalance, which can reduce GPU
efficiency and make CPU execution competitive~\cite{irregular}. As a result, the
fastest configuration may not be limited to CPU-only or GPU-only
execution; intermediate CPU-GPU partitions may also be efficient
depending on the input matrix and platform. This makes sparse CG
suitable for studying whether integrated CPU-GPU platforms such
as GH200 broaden the set of useful coscheduled execution points.

\subsection{Coscheduled CG Structure}
\label{sec:design-cg-algorithm}

\begin{algorithm}[t]
\small
\caption{Coscheduled CPU--GPU conjugate gradient solver}
\label{alg:hybrid-cg}
\begin{algorithmic}[1]
\Require \(A\), \(b=\mathbf{1}\), \(x_0=\mathbf{0}\),
         \(\tau=10^{-8}\), \(I_{\max}=2000\)
\State \(x\gets x_0\), \(r\gets b-Ax\), \(p\gets r\),
       \(\rho\gets r^Tr\)
\For{\(k=0,\ldots,I_{\max}-1\)}
    \If{\(\sqrt{\rho}<\tau\)}
        \State \textbf{break}
    \EndIf
    \State CPU and GPU compute their partitions of \(q\gets Ap\)
    \State CPU and GPU compute partial contributions to \(p^Tq\)
    \State Synchronize, reduce \(p^Tq\), and compute
           \(\alpha\gets\rho/(p^Tq)\)
    \State CPU and GPU update their partitions of
           \(x\gets x+\alpha p\) and \(r\gets r-\alpha q\)
    \State \(\rho_{\mathrm{old}}\gets\rho\)
    \State CPU and GPU compute partial contributions to \(r^Tr\)
    \State Synchronize and reduce \(\rho\gets r^Tr\)
    \State \(\beta\gets\rho/\rho_{\mathrm{old}}\)
    \State CPU and GPU update their partitions of \(p\gets r+\beta p\)
    \State Make the updated \(p\) visible according to the memory mode
\EndFor
\end{algorithmic}
\end{algorithm}

Algorithm~\ref{alg:hybrid-cg} summarizes the coscheduled CG iteration. Each
iteration contains three phases: sparse matrix-vector multiplication, solution
and residual update, and search-direction update. The work is split between the
CPU and GPU based on matrix rows. The CPU and GPU execute their assigned row
partitions concurrently in each phase, and scalar reductions are synchronized
between phases.

All matrix values, vectors, reductions, and CG coefficients use
double-precision floating-point arithmetic. We use unpreconditioned CG. 
The search-direction vector \(p\) is the main shared vector in the solver.
Although each processor updates only its own entries of \(p\), the next SpMV
may read any entry through the sparse column indices:
\[
q_i=\sum_{j\in A(i)} A_{ij}p_j,
\]

Therefore, the complete updated \(p\) vector must be visible to both CPU and
GPU partitions before the next iteration. This requirement is the main point
where the memory-management strategy affects the implementation.

The CPU and GPU use different reduction orders. CPU partial sums depend on the
OpenMP row partition, while GPU partial sums are formed using block-level
reductions and combined across blocks. These differences can produce small
floating-point variations in \(p^Tq\), \(r^Tr\), and the final residual. 
Therefore, for
each configuration, we record the iteration count, convergence
status, and final residual. After each solve, we recompute the residual
directly from the final solution to verify numerical correctness. 

\subsection{Data Layout and Work Partitioning}
\label{sec:design-partitioning}

Input matrices are read in Matrix Market format and converted to compressed
sparse row (CSR) format. For symmetric inputs, off-diagonal entries are
expanded so that the solver operates on the full sparse matrix. The CSR arrays
are copied to device memory during initialization for GPU-enabled runs, while
the CPU retains the host CSR representation. Since \(A\) is read-only, the
matrix data is not exchanged during the iterative solve. Work is partitioned by rows. Given a matrix with \(n\) rows and GPU fraction
\vspace{-2pt}
\(g\), the first
\[
n_{\mathrm{gpu}}=\operatorname{round}(g\cdot n)
\]

rows are assigned to the GPU, and the remaining rows are assigned to the CPU
threads. Thus, \(g=0\) represents a CPU-only execution, \(g=1\) represents
a GPU-only execution, and intermediate values represent hybrid CPU--GPU
coscheduling. Because sparse matrices can have nonuniform row lengths, equal
row partitions do not necessarily imply equal work. 
Hence, we record both
the number of rows and nonzeros assigned to each side.

\subsection{CUDA and OpenMP Execution}
\label{sec:design-kernels}

The solver uses an OpenMP parallel region to coordinate CPU and GPU execution.
In hybrid configurations, one OpenMP thread launches CUDA kernels and manages
GPU-side execution, while the remaining OpenMP threads process CPU-owned rows
by dividing the rows equally among them. The CG application is run with
32 OpenMP threads. 

The SpMV and dot product are fused into one GPU kernel, and the update of
\(x\), \(r\), and \(r^Tr\) is also fused. These fused kernels reduce extra
kernel launches and avoid additional passes over intermediate vectors.
Barriers and CUDA synchronization points are retained where reductions or
shared-vector visibility are required, as these costs are part of
coscheduled execution.

\subsection{Memory Execution Modes}
\label{sec:design-memory-modes}

The three execution modes mainly differ in how the shared search-direction
vector \(p\) is made visible to the CPU and GPU. In explicit-copy execution,
separate host and device copies of \(p\) are maintained; after each iteration,
the GPU-owned segment is copied to the host and the CPU-owned segment is copied
to the device. In managed-memory execution, the CG vectors use CUDA managed
memory, allowing \(p\) to be shared without explicit application-level copies.
In mapped-memory execution, \(p\) resides in page-locked host memory and is
accessed by the GPU through a device pointer. Although managed and mapped memory
remove code for explicit \(p\)-exchange, their costs may appear as migration,
remote access, coherence, kernel, or synchronization overhead.

\section{Evaluation}
\label{sec:evaluation}



Our evaluation studies how the NVIDIA GH200 
platform changes the
coscheduling space of CG compared to
the discrete H100 PCIe platform. We define the space as the set
of CPU-GPU work splits and memory-management strategies that produce
competitive run times for a given matrix.

The evaluation is guided by the following research questions:

\begin{itemize}
    \item \textbf{RQ1: Integrated versus discrete execution.}
    How do GH200 and H100 PCIe differ across CPU-GPU work splits
and memory-management strategies? In particular, does
GH200 make intermediate CPU-GPU splits more practical
than on the discrete H100 PCIe platform?

    \item \textbf{RQ2: Memory-management tradeoffs.}
    What is the runtime cost of managed memory and mapped memory relative
to explicit-copy execution, and how does this cost differ
between GH200 and H100 PCIe?

    \item \textbf{RQ3: Interconnect-bandwidth bound.}
    Can the H100
PCIe performance gap be explained by explicit transfer
bandwidth alone?
\end{itemize}

Table~\ref{tab:platforms} summarizes the  two platforms
used in the experiments. Both implementations are compiled for the Hopper architecture using
\texttt{-arch=sm\_90}. We use the same source code and compiler options on
both systems:
\texttt{-O3 -std=c++17 -Xcompiler -fopenmp}.

\begin{table}[tb]
\caption{Specifications of the evaluated platforms. Bandwidth values are peak specifications by the vendor.}
\label{tab:platforms}
\small
\setlength{\tabcolsep}{4pt}
\renewcommand{\arraystretch}{0.95}
\begin{tabularx}{\columnwidth}{
  l
  >{\raggedright\arraybackslash}X
  >{\raggedright\arraybackslash}X
}
\toprule
\textbf{Component} & \textbf{GH200} & \textbf{H100} \\
\midrule

CPU
& NVIDIA Grace, 72 Arm Neoverse-V2 cores
& \(2\times\) AMD EPYC 9454, 96 cores total \\

Host memory
& 480~GB LPDDR5X, 500~GB/s
& 1.5~TiB DDR5, 921.6~GB/s aggregate \\

GPU
& GH200 Hopper, 96~GB HBM3, 4.0~TB/s
& H100, 94~GB HBM3, 3.35~TB/s \\

CPU--GPU link
& NVLink-C2C, 900~GB/s aggregate
& PCIe Gen5 x16, 128~GB/s aggregate \\

Software
& CUDA 13.0, driver 580.65.06,
  \texttt{nvcc} 13.0.88
& CUDA 13.1, driver 610.43.02,
  \texttt{nvcc} 13.1.80 \\

\bottomrule
\end{tabularx}
\end{table}

\subsection{Workloads and Configurations}
\label{sec:evaluation-configurations}

The benchmark uses eight sparse matrices from the SuiteSparse Matrix
Collection~\cite{davis2011suitesparse}, covering 63,838 to 1,498,023 rows,
4,817,870 to 77,651,847 nonzeros, and approximately 5.0 to 398.8 nonzeros per
row. This range provides different sparsity structures and row-length
distributions, for which equal row partitions may not correspond to equal
computational work.

For each matrix and platform, we evaluate the cross-product of CPU-GPU work
division and memory-management strategy. The primary GPU row fractions are
\(g \in \{0.00,0.25,0.50,0.75,1.00\}\). To check whether the coarser sweep missed an optimum near the
GPU-only endpoint, we also evaluated \(g \in \{0.80,0.85,0.90,0.95\}\).
These additional points did not change the best split identified for any
matrix or platform. Each split is evaluated under explicit copy, managed
memory, and mapped memory. Each configuration uses three warm-up passes followed by ten measured passes. 
For each measured pass, we record solve time, iteration count, convergence
status, residuals, CPU compute time, GPU task time, barrier time, and copy
times. 

Explicit-copy runs additionally separate device-to-host and
host-to-device \(p\)-exchange time. 
To verify that our custom GPU implementation was reasonably optimized, we compared its isolated CSR SpMV kernel against cuSPARSE. Using 256-thread blocks, the custom kernel achieved \(72.7\%\) of the cuSPARSE performance, including cuSPARSE setup overhead, with \(49.86\%\) mean DRAM utilization and \(2.00\)~TB/s measured DRAM bandwidth, compared to \(56.69\%\) and \(2.28\)~TB/s for cuSPARSE.

Configurations are
classified as competitive when their run time is within \(1.25\times\) of the best run time for the same matrix and platform.
All evaluated configurations reached the maximum iteration count without satisfying the convergence tolerance. Therefore, the reported run times represent
fixed-work performance for 2000 CG iterations rather than time to
convergence.

\subsection{RQ1: Integrated vs. Discrete Coscheduling}
\label{sec:rq1}

\begin{figure*}[t]
    \centering
    \includegraphics[width=\textwidth]
        {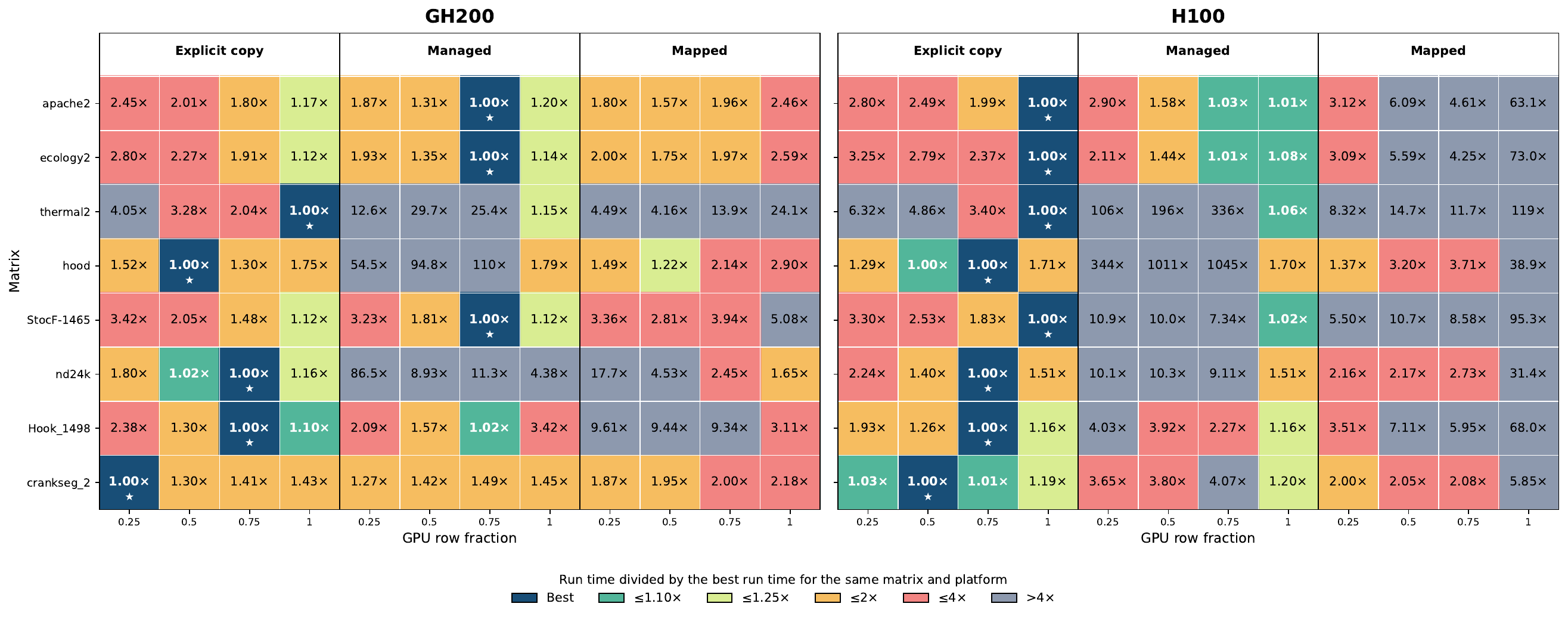}
    \caption{Normalized CG run time landscape on GH200 and H100 PCIe Platform. Each cell
    reports slowdown relative to the fastest configuration for the same matrix
    and platform; \(1.00\times\) marks the optimum. Lower is better.}
    \label{fig:rq1-landscape}
\end{figure*}

RQ1 compares the configuration landscapes of the integrated GH200 and discrete
H100 PCIe platforms. Figure~\ref{fig:rq1-landscape} reports each configuration
as a \textbf{\textit{slowdown}} relative to the fastest configuration for the same matrix and
platform. Thus, \(1.00\times\) marks the best observed configuration, while
larger values indicate increasing sensitivity to GPU fraction and
memory-management mode.

The two platforms favor different configurations. On GH200, hybrid execution is optimal for 6 of 8 matrices. Across all eight
matrices, managed memory provides the overall best result for 3 matrices and
explicit copy for 5. On H100, all optima use explicit copy; 3 matrices favor hybrid execution, 4 favor GPU-only execution, and 1 favors CPU-only execution.

Table~\ref{tab:absolute-runtime-summary} includes absolute run times. For each platform, it reports the fastest CPU-only
and GPU-only configurations and the fastest hybrid configuration time, with the corresponding hybrid fraction shown in
the \(g\) column.

\begin{table}[t]
\centering
\caption{Mean CG run times (s) by execution category. Bold marks the
fastest category for each matrix and platform.}
\label{tab:absolute-runtime-summary}

\resizebox{\columnwidth}{!}{%
\begin{tabular}{@{}l r r r c | r r r c@{}}
\toprule
& \multicolumn{4}{c|}{\textbf{GH200}}
& \multicolumn{4}{c}{\textbf{H100}} \\
\cmidrule(lr){2-5}
\cmidrule(lr){6-9}
\textbf{Matrix}
& \textbf{CPU}
& \textbf{GPU}
& \textbf{Hyb.}
& \textbf{\(g\)}
& \textbf{CPU}
& \textbf{GPU}
& \textbf{Hyb.}
& \textbf{\(g\)} \\
\midrule

\texttt{Hook\_1498}
& 4.978 & 2.064 & \textbf{1.881} & 0.75
& 5.649 & 3.049 & \textbf{2.640} & 0.75 \\

\texttt{StocF-1465}
& 2.106 & 0.556 & \textbf{0.499} & 0.75
& 2.216 & \textbf{0.801} & 1.469 & 0.75 \\

\texttt{apache2}
& 0.360 & 0.199 & \textbf{0.170} & 0.75
& 0.566 & \textbf{0.311} & 0.322 & 0.75 \\

\texttt{crankseg\_2}
& \textbf{1.079} & 2.178 & 1.887 & 0.25
& \textbf{1.337} & 2.405 & 2.015 & 0.50 \\

\texttt{ecology2}
& 0.463 & 0.209 & \textbf{0.187} & 0.75
& 0.653 & \textbf{0.335} & 0.340 & 0.75 \\

\texttt{hood}
& 0.563 & 0.566 & \textbf{0.324} & 0.50
& 0.795 & 0.915 & \textbf{0.537} & 0.75 \\

\texttt{nd24k}
& 1.718 & 0.809 & \textbf{0.696} & 0.75
& 2.042 & 1.008 & \textbf{0.667} & 0.75 \\

\texttt{thermal2}
& 1.035 & \textbf{0.307} & 0.626 & 0.75
& 2.669 & \textbf{0.443} & 1.504 & 0.75 \\

\bottomrule
\end{tabular}%
}
\end{table}

The absolute results confirm that no single execution category dominates all
workloads. The CPU-only optimum observed for
\texttt{crankseg\_2} on both systems also shows that GPU participation may add
more overhead than benefit for some sparse structures.

Overall, GH200 broadens the practical sparse CG coscheduling space by making
hybrid execution optimal more frequently and allowing managed memory to
achieve the best run time for several workloads. The H100 platform retains useful hybrid
configurations, but its optima are more strongly concentrated around
explicit-copy and GPU-dominant execution. The preferred configuration therefore
depends on both the sparse matrix and the target platform.

\subsection{RQ2: Memory-Management Tradeoffs}
\label{sec:rq2}

\begin{figure}[tb]
    \centering
    \includegraphics[width=\columnwidth]
        {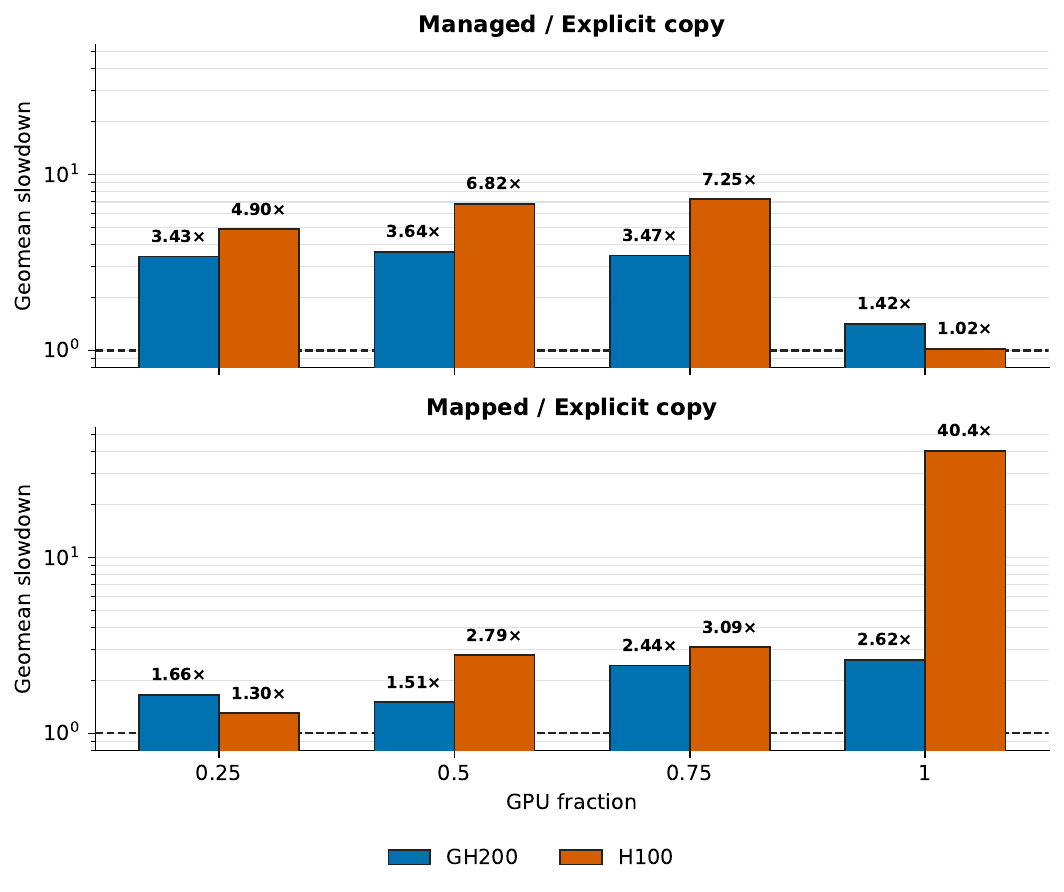}
    \caption{Geomean run time penalty of managed and mapped memory
    relative to explicit copy at the same GPU fraction. The dashed line marks
    parity. Lower is better.}
    \label{fig:memory-penalty}
    \vspace{-10pt}
\end{figure}

RQ2 compares managed and mapped memory with explicit-copy execution at the same
GPU fraction. Figure~\ref{fig:memory-penalty} shows the geomean run time
ratio across matrices, where \(1.00\times\) denotes parity with explicit copy.
Matrix-specific behaviour is shown in Figure~\ref{fig:rq1-landscape}.

Managed-memory performance depends strongly on GPU fraction and matrix. At
hybrid splits, GH200 incurs a smaller aggregate penalty than the H100 PCIe
platform, indicating that its coherent CPU--GPU memory system reduces, but does
not eliminate, managed-memory overhead when both processors participate.

At \(g=1.00\), however, H100 managed memory is nearly equivalent to explicit
copy, with a geometric-mean slowdown of \(1.02\times\), compared with
\(1.32\times\) on GH200. One possible explanation is that GPU-only execution on
H100 allows managed pages to migrate to and remain in local HBM, whereas on
GH200 some pages may remain in Grace memory or incur placement and translation
overheads. Because we do not directly measure page residency or migration, this
remains a hypothesis.

Managed memory is within \(1.25\times\) of the best run time for 5 of 8 matrices
on GH200 and 6 of 8 on H100, but it is the best configuration for only 3
matrices on GH200. Thus, the large hybrid penalties are concentrated in a small
number of severe outliers. In particular, \texttt{thermal2} and
\texttt{hood} exhibit substantial increases in CPU, GPU, and synchronization
time during hybrid execution.

Profiling \texttt{hood} on GH200 links this behavior to repeated
unified-memory movement involving the shared vector \(p\). Nsight Systems records approximately \(1.87\)
million Unified Host-to-Device and \(2.45\) million Unified Device-to-Host
operations over 2000 iterations, while the average SpMV kernel time increases
from \(0.122\)~ms with explicit copy to \(21.86\)~ms with managed memory. By
contrast, \texttt{apache2}, for which managed memory performs well, generates
only 760 Unified Host-to-Device and 184 Unified Device-to-Host operations.
These results indicate that matrices with greater cross-partition access to
\(p\) can trigger repeated bidirectional unified-memory movement and stall both
processors, whereas locality-preserving access patterns incur little managed-memory overhead.

Mapped memory is less robust as GPU participation increases because the GPU
repeatedly accesses host-resident data. This is especially costly on the
PCIe-attached H100 platform.

Overall, GH200 makes managed memory more practical at hybrid CPU--GPU splits,
while H100 performs better under GPU-only managed execution. Mapped memory
remains highly sensitive to matrix access pattern, GPU fraction, and
interconnect, and is particularly expensive for GPU-dominant execution on the
discrete platform.

\subsection{RQ3: Interconnect-Bound Sensitivity}
\label{sec:rq3}

RQ3 estimates how much of the GH200--H100 performance difference can be
explained by CPU--GPU transfer bandwidth. Let \(T\) be the measured H100
explicit-copy run time and \(T_{\mathrm{transfer}}\) the sum of initial, final,
and per-iteration vector-transfer time. We scale only this component by the
peak aggregate bandwidth ratio
\(R=B_{\mathrm{GH200}}/B_{\mathrm{H100}}=900/128\approx7.03\):
\[
T_{\mathrm{bound}}
=
T-T_{\mathrm{transfer}}
+\frac{T_{\mathrm{transfer}}}{R}.
\]
CPU, GPU, synchronization, and other non-transfer costs remain unchanged.

\begin{figure}[tb]
    \centering
    \includegraphics[width=\columnwidth]
        {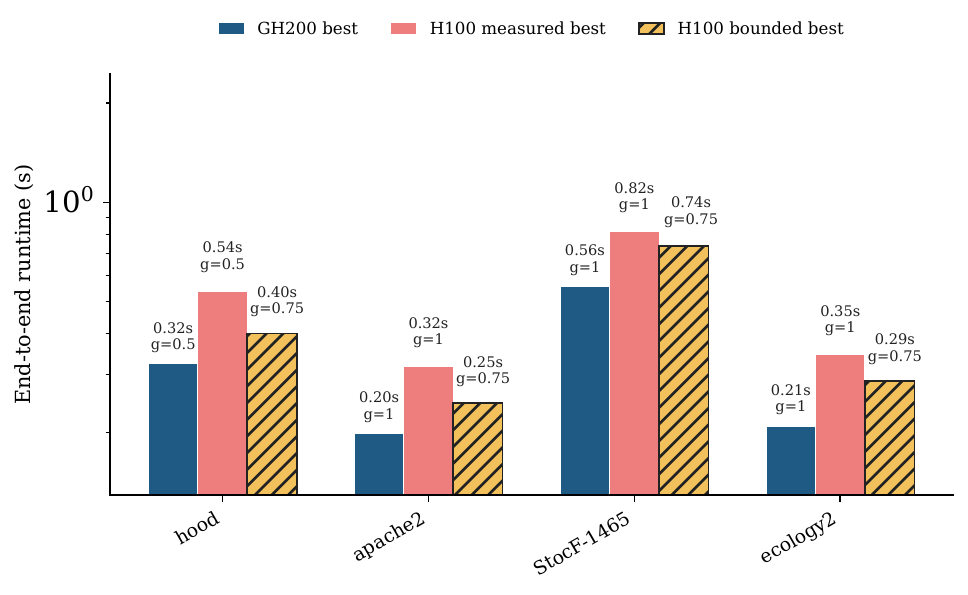}
    \caption{Measured GH200 and H100 explicit-copy run times and the H100
    transfer-bandwidth bound. Only cases whose preferred H100 split changes
    after scaling \(T_{\mathrm{transfer}}\) are shown.}
    \label{fig:rq3-transfer-bound}
    \vspace{-13.5pt}
\end{figure}

Figure~\ref{fig:rq3-transfer-bound} considers only explicit-copy configurations and shows the cases in which the preferred
H100 work split changes after applying the transfer-bandwidth bound. In each
case, the optimum shifts toward \(g=0.75\), indicating that lower communication
cost can make hybrid execution more favorable than GPU-only or more evenly
partitioned configurations.

The improvement remains matrix-dependent because scaling transfer time cannot
remove CPU, GPU, or synchronization bottlenecks. Even under this optimistic
bound, H100 does not consistently match GH200, showing that bandwidth explains only part of the observed system-level difference.

The analysis assumes ideal inverse scaling with peak bandwidth and does not
model latency, protocol overhead, transfer granularity, PCIe efficiency,
coherence, page placement, migration, or synchronization effects. Because the
platforms also differ in CPU architecture, host memory, GPU configuration, and
software environment, the result should be interpreted as an upper-bound
rather than a prediction of H100 performance with
NVLink-C2C.

Overall, CPU--GPU transfer bandwidth can materially affect run time and the
preferred coscheduling split, but it is insufficient by itself to reproduce the
GH200 coscheduling space. The remaining difference reflects broader
communication, memory-system, and processor characteristics.
\section{Future Work}
\label{sec:future-work}

Future work can extend this study in four directions. First, the evaluation
should include workloads beyond sparse CG to determine whether the
observed GH200 trends generalize across different memory-access,
synchronization, and communication-to-computation characteristics.

Second, the fixed sweep should be replaced with automated selection of
CPU--GPU partitions and memory strategies. Runtime or autotuning approaches could explore finer-grained and
non-contiguous splits, with systems such as CoreTSAR~\cite{scogland2014coretsar}
providing a useful foundation.

Third, detailed profiling is needed to separate migration, remote-access,
coherence, synchronization, and kernel overheads in managed and mapped memory.
These measurements could support predictive selection among memory management strategies.

Finally, future studies should evaluate power and energy, since the fastest
configuration may not minimize energy-to-solution. 
\section{Conclusion}

We presented a preliminary system-level study of CPU--GPU coscheduling on an
integrated NVIDIA GH200 platform and a discrete H100 PCIe platform using sparse
CG. We evaluated workload partitioning and three memory-management strategies
across both systems.

The results show that GH200 supports a broader coscheduling space, with hybrid
CPU--GPU splits and managed-memory execution becoming competitive for more
matrices, whereas H100 PCIe more frequently favors GPU-dominant explicit-copy
configurations. Mapped memory is not consistently effective, particularly on
the discrete platform.

Our optimistic transfer-bandwidth analysis indicates that PCIe bandwidth alone
does not explain the observed system-level differences. CPU architecture,
memory-system behavior, synchronization, CPU-side execution, GPU configuration,
and software environment also contribute. Overall, integrated CPU--GPU
platforms can make coscheduling more effective by expanding the set of
competitive workload partitions and memory-management choices.

\begin{acks}
This work was supported in part by the Synergistic Environments for
Experimental Computing (SEEC) Center at Virginia Tech.
\end{acks}

\bibliographystyle{ACM-Reference-Format}
\bibliography{paper}

\end{document}